# Advanced SERS sensor based on capillarity-assisted preconcentration through gold nanoparticles-decorated porous nanorods

Longjian Xue, [§,‡,*] Wei Xie, [ǁ,&] Leonie Driessen, [#] Katrin F. Domke, [#] Yong Wang, [%,*] Sebastian Schlücker, [&] Stanislav N. Gorb, [$] and Martin Steinhart[‡,*]

[§] Department of Materials Engineering, School of Power and Mechanical Engineering, Wuhan University, South Donghu Road 8, 430072 Wuhan, China

[‡] Institut für Chemie neuer Materialien, Universität Osnabrück, Barbarastr. 7, 49069 Osnabrück, Germany

[ǁ] Key Laboratory of Advanced Energy Materials Chemistry (Ministry of Education), College of Chemistry, Nankai University, Tianjin 300071, China

[$] Zoologisches Institut, Christian-Albrechts-Universität zu Kiel, Am Botanischen Garten 1-9, 24098 Kiel, Germany

[&] Fakultät für Chemie, Universität of Duisburg-Essen, Germany

[#] Max Planck Institute for Polymer Research; Ackermannweg 10 - D-55128 Mainz.

[%] State Key Laboratory of Materials-Oriented Chemical Engineering and College of Chemistry and Chemical Engineering, Nanjing University of Technology, China

* To whom correspondence should be addressed. E-mail: xuelongjian@whu.edu.cn, yongwang@njut.edu.cn, martin.steinhart@uni-osnabrueck.de

**Abstract**
A preconcentrating surface-enhanced Raman scattering (SERS) sensor for the analysis of liquid-soaked tissue, tiny liquid droplets and thin liquid films without the necessity to collect the analyte is reported. The SERS sensor is based on a blockcopolymer membrane containing a spongy-continuous pore system. The sensor's upper side is an array of porous nanorods having tips functionalized with Au nanoparticles. Capillarity in combination with directional evaporation drives the analyte solution in contact with the flat yet nanoporous underside of the SERS sensor through the continuous nanopore system toward the nanorod tips where non-volatile components of the analyte solution precipitate at the Au nanoparticles. The nanorod architecture increases the sensor surface in the detection volume and facilitates analyte preconcentration driven by directional solvent evaporation. The model analyte 5,5′-dithiobis(2-nitrobenzoic acid) can be detected in a $1 \times 10^{-3}$ m solution ≈300 ms after the sensor is brought into contact with the solution. Moreover, a sensitivity of 0.1 ppm for the detection of the dissolved model analyte is achieved.

**Introduction**

Surface-enhanced Raman scattering (SERS)[1,2] has been explored as an analytical method for trace analysis and *in situ* monitoring of molecular processes. The SERS effect is based on local field enhancement related to the occurrence of surface plasmons at the surfaces of metal substrates. Thus, analytes located in close vicinity to the surfaces of SERS-active metals, such as gold, show dramatically enhanced Raman scattering intensities and can be detected with significantly improved sensitivity. Previous research has mainly focused on the design of SERS substrates with nanostructured surfaces exhibiting high densities of so-called hot spots (spots characterized by extraordinary field enhancement) to enhance Raman scattering intensities as much as possible.[3–7] For example, arrays of nanorods and nanotubes coated with metal or decorated with metal nanoparticles have been explored as highly efficient SERS substrates.[8–12] Recent research has focused on utilizing SERS for *in situ* high-throughput detection of analytes as well as for the design of opto-fluidic devices combining microfluidics and SERS.[13–15] "Flow-through" microfluidic devices[16] or photonic crystal fibers[17] have been tested as potential device architectures to this end.

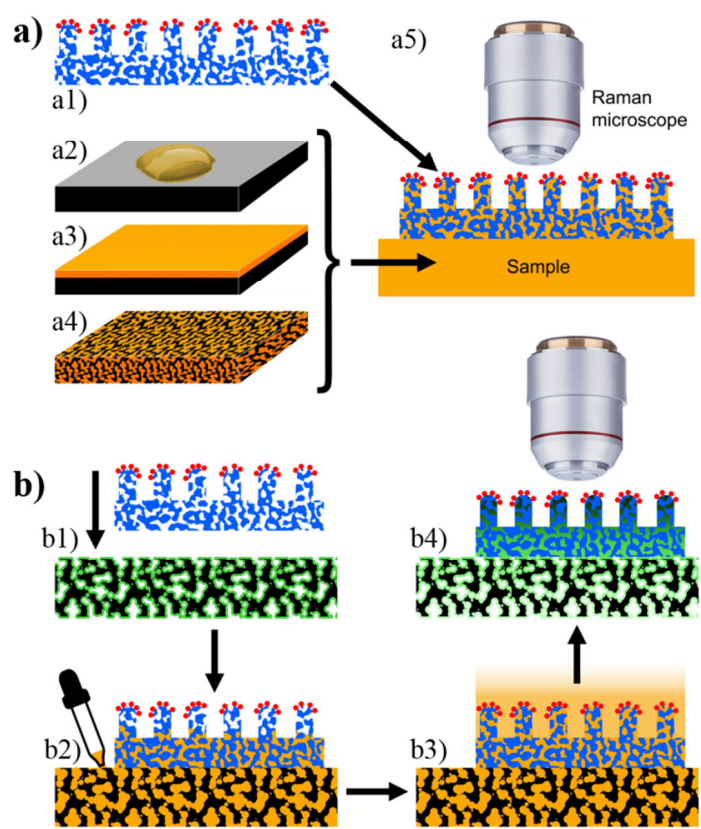

**Figure 1.** Application of the capillarity-based preconcentrating SERS sensor. a1) The SERS sensor consists of a porous BCP membrane (blue) with an array of porous BCP nanorods at its upper surface. The nanorods are functionalized with AuNPs (red). The smooth porous underside of the SERS sensor is brought into contact with a surface covered by a2) a droplet or by a3) a thin film of the analyte solution (orange) or with a4) tissue soaked with analyte solution. a5) Capillarity and directional solvent evaporation drive the analyte solution to the tips of the porous nanorods decorated with AuNPs as SERS probes, which are located in the detection volume of a Raman microscope. The analyte preconcentrates at the AuNPs in the detection volume of the Raman microscope. b1)–b4) Possible application in which (b1) the SERS sensor is placed on tissue containing a solid contamination (green). (b2) An injectedsolvent (orange) dissolves the contamination. (b3) Directional evaporation of the solvent toward the nanorod tips of the SERS sensor results in (b4) preconcentration of the analyte at the AuNPs at the nanorod tips for SERS-supported detection.

Here, we report a capillarity-based preconcentrating SERS sensor (Figure 1a) for the detection of analytes contained in small pieces of tissue, paper, and membranes as well as in liquid droplets or in thin liquid films without the necessity to collect the analyte. The SERS sensor architecture is based on a block copolymer (BCP) membrane containing a continuous spongy nanopore system (Figure 1a1). The flat yet nanoporous underside of the SERS sensor is brought into contact with samples such as solid substrates covered by tiny droplets (Figure 1a2) or thin liquid films (Figure 1a3) of an analyte solution. Also, tissue soaked with analyte solution (Figure 1a4) can be investigated in this way. Capillarity in combination with directional evaporation drives the analyte solution through the continuous nanopore system toward the upper air surface of the SERS sensor. The upper air surface of the SERS sensor consists of an array of porous nanorods having tips functionalized with gold nanoparticles (AuNPs) (Figure 1a1). The AuNPs employed as SERS probes are located in the detection volume of a Raman microscope (Figure 1a5). The presence of an array of porous nanorods increases the surface of the SERS sensor within the detection volume of the Raman microscope; thus, more AuNPs are located in the detection volume than in the case of a smooth upper SERS sensor air surface. Moreover, the large specific surface area of the porous nanorod array supports directional evaporation of analyte solutions so that more analyte solution is driven into the SERS sensor. Evaporation-induced depletion of the solvent results in precipitation and preconcentration of nonvolatile analytes at the AuNP-decorated tips of the porous nanorods. Conventional SERS samples, which are prepared by dropping analyte solution on a nanostructured metal surface or by mixing the analyte solution with metal nanoparticle suspensions, suffer from poor reproducibility because of inhomogeneous spatial distribution of the analyte and/or the aggregation of the metal nanoparticles. The SERS sensor proposed here, taking advantage of directional evaporation and enrichment of the analyte at the AuNPs, protects the SERS substrate from any destructive treatment and ensures homogeneous spatial distribution of the analyte on the SERS sensor. Therefore, the poor reproducibility of conventional SERS detection can be overcome. Moreover, integrating the SERS sensor architecture reported here into microfluidic devices may enable online *in situ* detection of reaction intermediates, which would not be possible with conventional flat SERS substrates, such as SERStrate. Furthermore, a careful design of the pore wall chemistry and/or the pore size might even be used to separate components of multicomponent analyte solutions for online monitoring.

In order to prove the concept, a possible application of the SERS sensor, for which conventional flat SERS substrates are not suitable, is displayed in Figure 1b. Dry tissue containing analyte is brought into contact with the SERS sensor (Figure 1b1). A solvent for the analyte is injected into the tissue (Figure 1b2). Directional evaporation of the injected solvent drives the analyte toward the tips of the porous nanorods of the SERS sensor where the analyte preconcentrates at the AuNPs (Figure 1b3). After solvent evaporation, the analyte can be identified with a Raman microscope exploiting SERS (Figure 1b4). The combination of sample and SERS sensor can be flexibly handled. The reliability of the SERS sensing benefits from the possibility to independently acquire several SERS spectra at different spots even within a small sample area. If a matrix of 3 × 3 Raman spectra per sample is taken at different spots on the sample surface with a lateral step width of 1 μm using a state-of-the-art Raman microscope, the sample area required for an analysis would be as small as 9 μm$^2$.

The SERS sensor with an overall thickness of ≈110 μm consisted of the diblock copolymer polystyrene-*b*-poly(2-vinyl pyridine) (PS-*b*-P2VP) and was produced by adapting procedures reported previously.[18,19] In brief, the PS-*b*-P2VP was molded against self-ordered nanoporous anodic aluminium oxide (AAO), which was removed by etching (Figure 2a). Continuous pore systems were then obtained by selective swelling-induced pore generation.[20–22] While the underside of the SERS sensor to be brought into contact with the sample was porous but smooth, the upper air surface consisted of an array of porous PS-*b*-P2VP nanorods pointing toward the Raman microscope. The nanoporous PS-*b*-P2VP nanorods had a diameter of ≈300 nm and a length of ≈1.5 µm (Figure 2b). The nearest-neighbor distance between the porous PS-*b*-P2VP nanorods amounted to ≈500 nm. Continuous sponge-like pore systems open to the environment with an average pore diameter of ≈98 nm and a specific surface of 45.5 m$^2$/g[18] penetrated the entire SERS sensor from its smooth underside to the tips of the porous PS-*b*-P2VP nanorods. The surfaces of the pore walls consisted of P2VP.[22] Pore size and pore morphology are, besides the conditions applied during swelling-induced pore generation, influenced by the volume ratios of the PS anf P2VP blocks and by the nanorod diameter.[22]

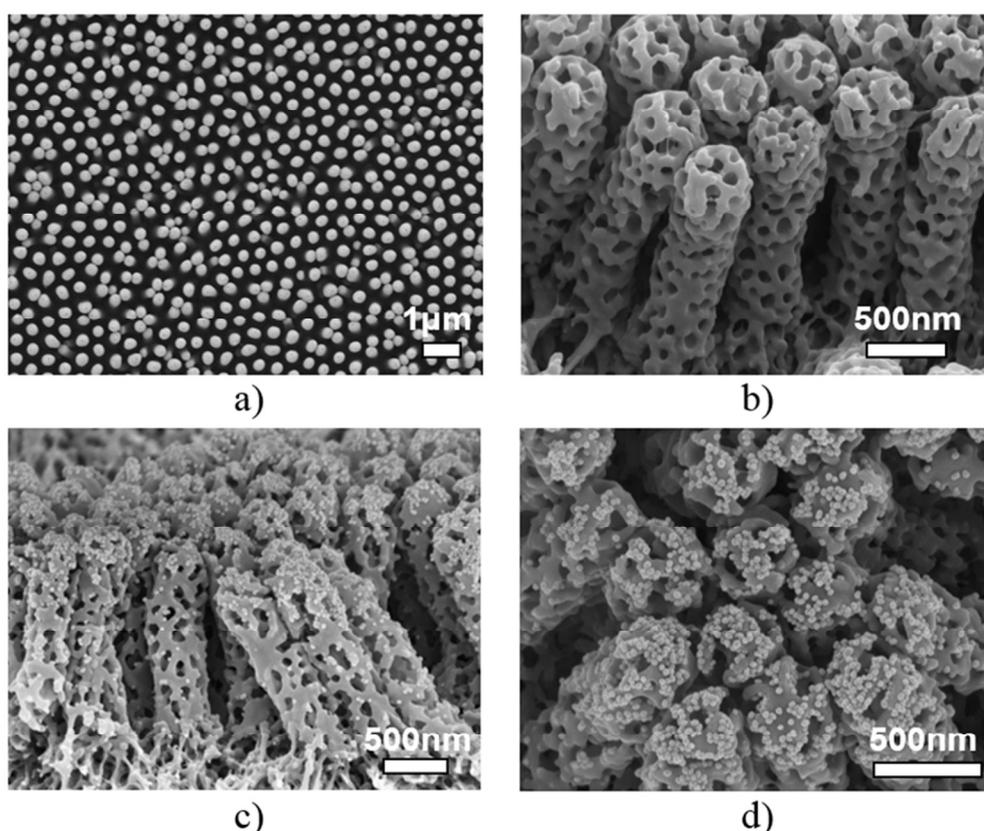

**Figure 2.** Scanning electron microscopy images of a) a PS-*b*-P2VP nanorod array released from self-ordered AAO and b) a porous PS-*b*-P2VP nanorod array obtained by swelling-induced pore generation. c) Side view and d) top view of a SERS sensor consisting of nanoporous PS-*b*-P2VP nanorods functionalized with AuNPs connected to a nanoporous PS-*b*-P2VP membrane.

Citrate-stabilized AuNPs with a mean diameter of ≈35.5 nm (Figure S1, Supporting Information) were attached to the tips of the porous PS-*b*-P2VP nanorods (Figure 2c,d) by floating the array of porous PS-*b*-P2VP nanorods on an acidic suspension of the AuNPs. The size of AuNPs was chosen according to the thickness of the pore walls (≈40 nm, Figure S2, Supporting Information) at the nanorod tips where the AuNPs attach to maximize the efficiency of the SERS detection using Raman microscopes. The pyridyl moieties of the P2VP blocks at the tips of porous PS-*b*-P2VP nanorods were partially protonated as P2VP is protonated at pH values smaller than ≈4.1.[23] Thus, besides van der Waals interactions also electrostatic interactions may support adhesion of the negatively charged AuNPs to the positively charged tips of the porous PS-*b*-P2VP nanorods. It should be noted that the AuNPs were not only attached directly to the nanorod tips but also away from the latter (Figure 2c). Since the focal volume of a Raman microscope typically has a depth of a few micrometers, more analyte precipitated at AuNPs is located within the focal volume so that the sensitivity of the SERS sensor is enhanced. After the attachment of the AuNPs, the porous PS-*b*-P2VP nanorods were partially condensed (Figure 2c,d). The agglomeration of the porous PS-*b*-P2VP nanorods might be advantageous for SERS because clustering may result in the formation of additional SERS hot spots between AuNPs attached to adjacent porous PS-*b*-P2VP nanorods.[24]

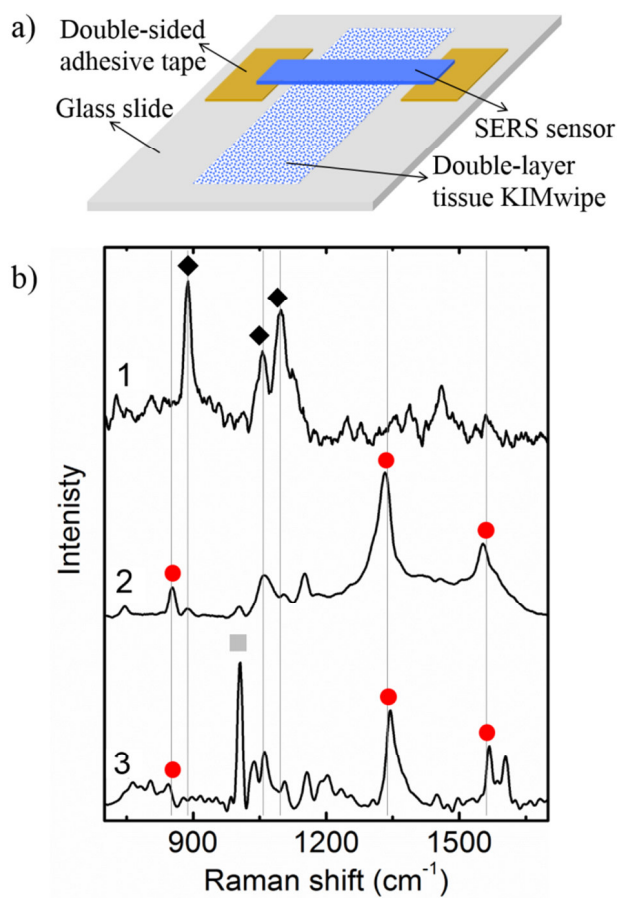

**Figure 3.** a) Schematic drawing of the configuration used for tests of the SERS sensor. b) (1) SERS spectrum of ethanol obtained by placing the SERS sensor on tissue soaked with ethanol. (2) SERS spectrum acquired ≈35 s after a 10 × 10$^{-3}$ m solution of DTNB in ethanol was dropped onto the uncovered part of the tissue underneath the SERS sensor using the configuration shown in (a). (3) Raman spectrum obtained by focusing on the tips of porous PS-b-P2VP nanorods not functionalized with AuNPs that had been immersed in a 10 × 10$^{-3}$ m solution of DTNB in ethanol for ≈60 s followed by drying under vacuum. Red circles denote DTNB peaks, gray squares denote P2VP peaks, and black diamonds denote ethanol peaks. The baselines of all spectra were subtracted.

Using the test design sketched in Figure 3a, SERS sensors were placed on tissue paper in such a way that the tips of the AuNP-coated porous PS-*b*-P2VP nanorods pointed away from the tissue paper as follows. Two pieces of double-sided adhesive tape were glued on a glass slide at a distance of 3–4 mm. In between, we placed a double layer of tissue (KIMwipe) with a length of ≈10 mm and a width of 3–4 mm. A SERS sensor with a length of ≈5 mm and a width of ≈1 mm was glued onto the upper sides of the double-sided adhesive tape stripes in such a way that the long edge of the SERS sensor was oriented normal to the long edge of the tissue. Moreover, the smooth porous underside of the SERS sensor was in contact with the tissue while the tips of the porous PS-*b*-P2VP nanorods of the SERS sensor pointed away from the tissue. In our experiments, about 10 µL analyte solution was dropped onto the area of the tissue not covered by the SERS sensor. The small amount of applied analyte solution ensured that analyte solution was exclusively transported to the AuNP-functionalitzed nanorod tips of the SERS sensor through the sensor's pore system driven by capillarity combined with directed solvent evaporation. Raman and SERS spectra were then acquired from the AuNP-functionalized tips of the porous PS-*b*-P2VP nanorods.

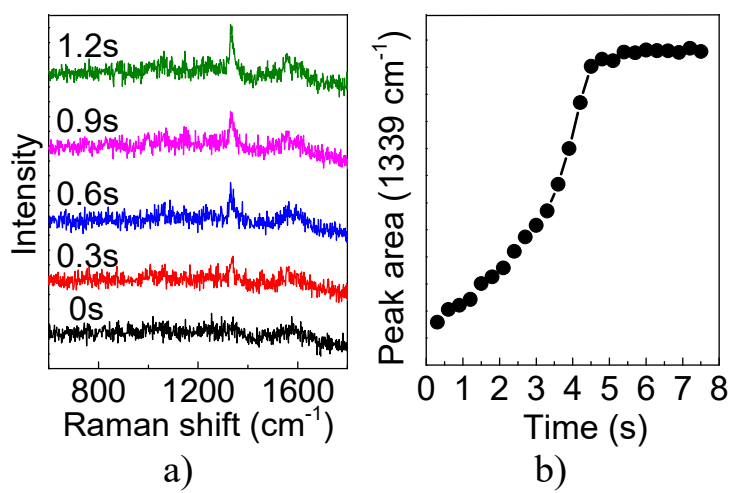

**Figure 4.** Temporal evolution of the SERS signals of DTNB. a) Spectra measured 0.0, 0.3, 0.6, 0.9, and 1.2 s after supply of $1 \times 10^{-3}$ m ethanolic DTNB solution onto the uncovered part of the tissue underneath the smooth porous underside of the SERS sensor using the configuration displayed in Figure 3a. b) Evolution of the areas of the DTNB peak at 1339 cm$^{-1}$ as function of the elapsed time.

In a first experiment, pure ethanol was dropped onto the tissue of the configuration displayed in Figure 3a. The ethanol immediately imbibed the SERS sensor. The SERS spectrum subsequently recorded (spectrum 1 in Figure 3b) shows peaks originating from ethanol at 887 cm$^{-1}$ (C-C stretching), 1056 cm$^{-1}$ (C-O stretching), and 1100 cm$^{-1}$ (CH$_3$ rock).[25] Raman spectra of pure ethanol and ethanol mixed with AuNPs in the absence of PS-*b*-P2VP are shown in Figure S3 (Supporting Information) for comparison. In a second experiment, a $10 \times 10^{-3}$ m solution of 5,5′-dithiobis(2-nitrobenzoic acid) (DTNB) in ethanol was dropped onto the tissue. After a wait time of 35 s we measured a Raman spectrum on the top surface of the SERS sensor (spectrum 2 in Figure 3b). The strongest signals in the thus-obtained spectrum at 855, at 1339, and at 1557 cm$^{-1}$ could be assigned to the nitro scissoring vibration, the symmetric nitro stretch, and the aromatic ring mode of DTNB.[26] These DTNB peaks with similar relative intensities also appeared in a SERS spectrum obtained from DTNB adsorbed on AuNPs in the absence of ethanol and of PS-*b*-P2VP (Figure S4, Supporting Information). However, no DTNB peaks appeared in a Raman spectrum of an ethanolic $10 \times 10^{-3}$ m DTNB solution (Figure S3, Supporting Information).

The strong DTNB signals appearing in spectrum 2 of Figure 3b were absent in the absence of AuNPs. We immersed a specimen consisting of porous PS-*b*-P2VP nanorods connected to an underlying ≈110 μm thick nanoporous PS-*b*-P2VP substrate into a $10 \times 10^{-3}$ m solution of DTNB in ethanol for ≈60 s, followed by drying under vacuum. A Raman spectrum was then recorded by focusing the laser beam on the tips of the porous PS-*b*-P2VP nanorods (spectrum 3 in Figure 3b). In this case the tips of the porous PS-*b*-P2VP nanorods were *not* functionalized with AuNPs so that SERS was not possible. In the Raman spectrum thus-obtained a relatively strong peak at 1007 cm$^{-1}$ originating from the symmetric ring breathing mode of P2VP[27] was observed, while only weak DTNB peaks at 1343 and 1570 cm$^{-1}$ appeared. The assignment of the peak at 1007 cm$^{-1}$ to P2VP was confirmed by the evaluation of Raman spectra of porous PS-*b*-P2VP nanorods and a blank SERS sensor (Figure S5, Supporting Information).

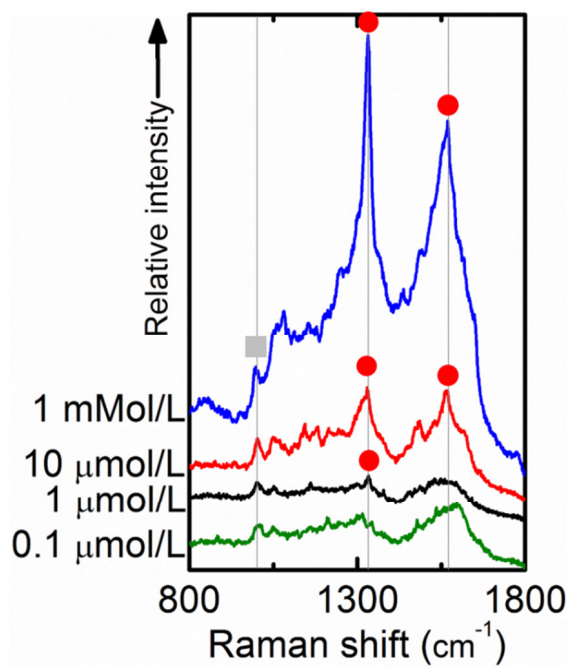

**Figure 5.** SERS signals of DTNB obtained with $0.1 \times 10^{-6}$, $1 \times 10^{-6}$, $10 \times 10^{-6}$, and $1 \times 10^{-3}$ m solutions of DTNB in ethanol using the configuration as displayed in Figure 3a. For each concentration, a new SERS sensor was used. The measurements were started immediately after the ethanolic DTNB solutions were dropped onto the uncovered part of the tissue underneath the SERS sensor. The measurements shown for each concentration are the average of ten consecutively measured spectra with an integration time of 1 s.

The area of the DTNB peak at 1339 cm$^{-1}$ in SERS spectrum (spectrum 2 in Figure 3b) was estimated to be ≈$6.1 \times 10^5$ times larger than in Raman spectrum (Figure S3, Supporting Information). Other striking differences between spectra 2 and 3 in Figure 3b are the different ratios of the relative peak intensities within the spectra. If the only difference between both spectra was the presence (spectrum 2) and the absence (spectrum 3) of the SERS effect, both spectra should nevertheless have shown identical fingerprints. However, relative to the P2VP peaks the DTNB peaks are much more pronounced in spectrum 2. This finding might be rationalized by assuming that efficient preconcentration of the low-molar mass DTNB molecules at the AuNPs occurred so that the SERS effect enhances the DTNB signals rather than the P2VP signals.

The robustness of the SERS sensor is related to the preconcentration of the analyte at the AuNPs used as SERS probes at the tips of the porous PS-*b*-P2VP nanorods. To monitor the preconcentration process, we dropped a $1 \times 10^{-3}$ m ethanolic DTNB solution onto the uncovered part of the tissue paper underneath the SERS sensor and started at the same time a series of consecutive SERS measurements with an integration time of 0.3 s (Figure 4a). Already in the second spectrum of the series taken after 0.3 s, a peak at ≈1339 cm$^{-1}$ originating from DTNB clearly appears. The area of the peak at ≈1339 cm$^{-1}$ increases with time during the successive measurements until a plateau was reached at 4.5 s (Figure 4b). A typical spectrum from the plateau showing strong DTNB signals recorded after 6 s is shown in Figure S6 (Supporting information). This outcome corroborates the notion that, as further DTNB solution is drawn to the AuNP-modified tips of the porous PS-*b*-P2VP nanorods where the solvent ethanol evaporates, DTNB is successively enriched at the surfaces of the AuNPs. The abrupt saturation of the peak area/time profile displayed in Figure 4b at 4.5 s likely indicates complete evaporation of ethanol. Therefore, the signal intensity could be even further enhanced if more pure solvent was applied to the tissue in order to push analyte remaining in the tissue and the interior of the sensor to the AuNPs at the sensor nanorods. In contrast, the SERS signal from a 1 × 10$^{-3}$ m DTNB solution in ethanol containing AuNPs remained constant within a time window of 10 min after mixing DTNB and AuNPs (Figure S7, Supporting Information), suggesting the absence of preconcentration under these conditions.

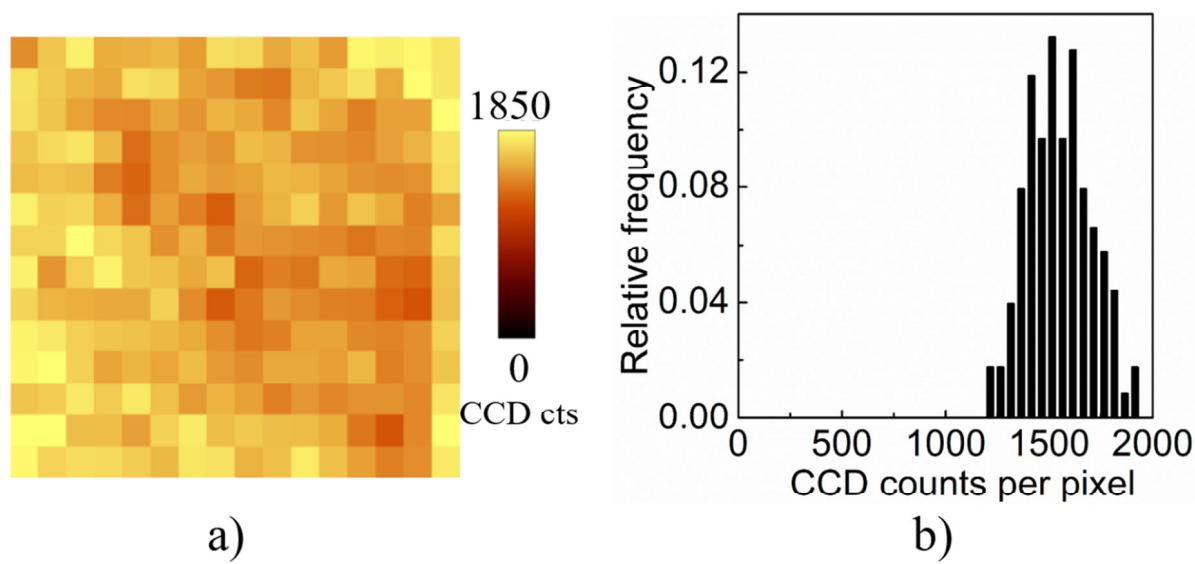

**Figure 6.** a) False-color mapping of a SERS sensor using the configuration as displayed in Figure 3a. The tissue was soaked with 10 × 10$^{-3}$ m ethanolic DTNB solution. After complete evaporation of the ethanol, we mapped the integral peak intensity of the symmetric nitro stretch of DTNB at 1339 cm$^{-1}$ in an area of 14 × 16 μm$^2$. b) Histogram to panel (a) displaying the relative density of the frequency of occurrence of the integral intensity values.

The sensitivity of DTNB detection was tested by dropping 0.1 × 10$^{-6}$, 1 × 10$^{-6}$, and 10 × 10$^{-6}$ m solutions of DTNB in ethanol onto uncovered parts of tissue underneath the SERS sensor (Figure 5). Immediately thereafter, ten successive Raman spectra with an integration time of 1 s were recorded and the intensities were averaged. The DTNB peak at ≈1339 cm$^{-1}$ clearly appeared in the curve obtained with a DTNB concentration of 1 × 10$^{-6}$ m. The spectrum obtained with a DTNB concentration of 10 × 10$^{-6}$ m shows strong DTNB peaks at ≈1339 and ≈1560 cm$^{-1}$. For a DTNB concentration of 0.1 × 10$^{-6}$ m, however, hardly any signal can be assigned to DTNB. However, longer wait times after supply of the analyte solution or pushing analyte remaining in the tissue and the interior of the sensor to the AuNPs at the sensor nanorods by further supply of pure solvent will allow analyte detection even for more diluted analyte solutions. Moreover, the detection sensitivity could be enhanced by optimizing the delivery of the analyte solution, the amount of analyte solution, and pore size as well as the chemistry of pore walls of the sensor. In our experimental design (Figure 3a), the surface area of the tissue is ≈9 times larger than than that of the SERS sensor. Thus, the solvent ethanol predominantly evaporates at the tissue surface. If the ethanol had exclusively evaporated at the sensor surface, more analyte would have been deposited at the AuNPs at the nanorod tips and the detection efficiency of the SERS sensor would have been even better.

Uniformity of the SERS response over sufficiently large areas is crucial for the usability and reliability of SERS sensors. We evaluated the uniformity of a SERS sensor using the test design as schematically displayed in Figure 3a. Then, 10 × 10$^{-3}$ m ethanolic DTNB solution was dropped onto the uncovered area of tissue located beneath the SERS sensor that then imbibed the SERS sensor. After the complete evaporation of the ethanol, we mapped the integrated intensity of the DTNB peak at 1339 cm$^{-1}$ originating from the symmetric nitro stretch over an area of 14 × 16 μm$^2$. Thus, we measured one spectrum per μm$^2$ and acquired spectra at 224 different spots. Figure 6a shows the color-coded intensity map. The histogram to Figure 6a seen in Figure 6b displays the relative frequency of occurrence of integrated peak intensity values. It is noticeable that the intensity map does not contain "cold spots" with low peak intensities, evidencing the pronounced uniformity of the SERS substrate. A relative standard deviation of 14.9% was estimated from five samples of three batches, indicating the good reproducibility of the sensor preparation.

A sensor for SERS detection of analytes contained in liquid drops or thin liquid films as well as in tissue, paper and membranes is reported. The SERS sensor consists of a porous block copolymer membrane, which can easily be prepared by selective swelling-induced pore generation. The smooth yet porous underside of the SERS sensor is brought into contact with the sample to be investigated. If the sample is covered by or contains analyte solution, the analyte solution imbibes the SERS sensor. If the analytes are solid, a suitable solvent for the analyte can be injected so that an analyte solution results. The air surface at the upper side of the SERS sensor consists of an array of porous block copolymer nanorods functionalized with AuNPs. A typical self-ordered AAO template with a diameter of 18 mm may yield up to 28 millions porous BCP membrane pieces with a size of 9 $\mu m^2$. The presence of the porous nanorod array at the air surface of the SERS sensor that is oriented toward the Raman microscope allows the immobilization of more AuNPs in the detection volume enabling more efficient SERS detection as compared to a smooth porous sensor surface. The porous nanorod array also supports the directional evaporation of the solvent. Directional evaporation and capillarity drive the analyte solution to the tips of the porous block copolymer nanorods located in the detection volume of a Raman microscope. Depletion of the solvent at the nanorod tips by directional evaporation results in preconcentration of the analytes at the tips of the porous nanorods functionalized with AuNPs. Thus, SERS-supported detection of the analytes with a Raman microscope is possible. Sample areas of a few $\mu m^2$ are sufficient for reliable analysis. As an example, DTNB could be detected after soaking macroscopic pieces of commercial tissue with 10 µL of a $1 \times 10^{-6}$ m solution of DTNB in ethanol. Using the same model configuration, DTNB could be detected after soaking the tissue with a $1 \times 10^{-3}$ m solution in ethanol in 300 µs. The performance of the SERS sensor could be optimized and the range of use cases could be extended by tailoring pore geometry, pore size, and pore wall chemistry of the sensor as well as by optimizing the supply of the analyte solution and the evaporation area of the sensor surface. We envision that the sensor configuration presented here could be used for forensic analytics or in arrays for high-throughput screenings.

**Experimental Section**

About 110 µm thick films consisting of the asymmetric diblock copolymer PS-*b*-P2VP with PS as the majority component ($M_n$ (PS) = 101 000 g mol$^{-1}$; $M_n$ (P2VP) = 29 000 g mol$^{-1}$; $M_w/M_n$ = 1.60; Polymer Source Inc., Canada) were prepared by dropping a solution of 100 mg PS-*b*-P2VP per mL tetrahydrofuran (THF) onto a silicon wafer. After the complete evaporation of THF, the PS-*b*-P2VP films were detached from the silicon wafers by immersing the samples in ethanol. Self-ordered AAO with a lattice period of ≈500 nm, a pore diameter of ≈300 nm and a pore depth of ≈1.5 µm[28] served as mold for the preparation of the nanorod arrays. The PS-*b*-P2VP films were placed on the AAO surface, which was heated to 220 °C, and kept a this temperature for 48 h, while a vacuum and a load of ≈160 mbar were applied. After cooling to room temperature at a rate of −1 K min−1, the PS-*b*-P2VP nanorod arrays were released from the AAO template by etching with a 40 wt% aqueous KOH solution, followed by rinsing with deionized water. The thus-obtained PS-*b*-P2VP nanorod arrays connected to PS-*b*-P2VP substrates (initially the PS-*b*-P2VP films located on the AAO surface) were faithful negative replicas of the AAO (Figure 2a). Arrays of PS-*b*-P2VP nanorods connected to PS-*b*-P2VP membranes (overall thickness ≈110 μm) that contained spongelike, continuous pores open to the environment reaching from the membrane underside to the tips of the nanorods (Figure 2b) were generated by selective swelling-induced pore formation in ethanol at 60 °C for 4 h.[18,19]

Tetrachloroauric (III) acid (HAuCl4), trisodium citrate, sodium borohydride and DTNB were purchased from Sigma-Aldrich and used without further purification. AuNPs were synthesized following procedures reported elsewhere.[29] 1 mL of an aqueous $3.88 \times 10^{-2}$ m citrate solution was added to 50 mL of a boiling solution of 0.01 wt% $HAuCl_4$ in water. The aqueous mixture having a pH value of 3.5 (measured using a S40 SevenMulti pH-meter, Mettler Toledo) was boiled for 20 min under vigorous stirring and then cooled to room temperature. The tips of porous PS-*b*-P2VP nanorods were coated with AuNPs by floating the porous PS-*b*-P2VP membranes on the thus-obtained aqueous AuNP suspension (nanorod tips pointed toward the suspension) for 1 h, following by three washing steps with deionized water. All Raman and SERS spectra were acquired using the 632.8 nm line of HeNe lasers for the excitation of Raman scattering. The Raman and SERS spectra shown in Figure 3 and Figures S3, S4 and S5 of the Supporting Information were measured using a Raman spectrometer from Princeton Instruments (Acton SP2500, 50 cm focal length, 1800 grooves $mm^{-1}$ grating). The diameter of the laser spot was ≈300 nm and the focal depth was ≈2.5 μm. The integration time per spectrum was 30 s. The Raman and SERS spectra shown in Figures 4 and 5 as well as in Figures S6 and S7 of the Supporting Information were acquired with a homemade setup described by Sabanés et al.[30] operated in the SERS mode. The laser power used amounted to 0.9 mW. The integration time per spectrum was 0.3 s (Figure 4; Figures S6 and S7 of the Supporting Information) and 1 s (Figure 5). The SERS mapping displayed in Figure 6 was performed with a WITec alpha 300R microspectrometer. The diameter of the laser spot was ≈300 nm, and the focal depth amounted to ≈2 μm. A grating monochromator (30 cm focal length, 600 grooves $mm^{-1}$ grating) equipped with a Peltier-cooled electron multiplying charge-coupled device camera was employed for recording the Raman spectra. The investigated sample was displaced in the *xy* plane. The integration time per spectrum was 2 s. The peak areas were determined without further fitting. Scanning electron microscopy investigations were carried out on a Zeiss Auriga microscope operated at an accelerating voltage of 3 kV. The samples were sputter-coated with a ≈5 nm thick iridium layer.

**Supporting Information**
Supporting Information is available from the Wiley Online Library or from the author.

**Acknowledgements**
The authors thank C. Hess and H. Tobergte for the preparation of AAO membranes. L.X. thanks the National Natural Science Foundation of China (51503156 and 51611530546) and the "Young 1000 talents" program for support. M.S. thanks the European Research Council (ERC-CoG-2014; project 646742 INCANA) for funding. Y.W., S.N.G., and M.S. thank the Alexander-von-Humboldt foundation for funding an institutional partnership in the "Research Group Linkage Programme." K.F.D. gratefully acknowledges generous support through the Emmy Noether program of the Deutsche Forschungsgemeinschaft (# DO 1691/1-1).

**Conflict of Interest**
The authors declare no conflict of interest.